\documentclass[aps,ams,amsmath,prl,twocolumn]{revtex4-1}
\usepackage{bm}
\usepackage{graphics}
\usepackage{graphicx}
\usepackage[dvipsnames]{xcolor}
\usepackage[colorlinks=true,linkcolor=blue,citecolor=BrickRed]{hyperref}
\begin{document}
\title{Macroscopic character of composite high temperature superconducting wires}

\author{S. A. Kivelson$^1$ and B. Spivak$^2$}
\affiliation{
1) Department of Physics, Stanford University, Stanford, California 94305, USA  \\
2) Department of Physics, University of Washington, Seattle, Washington 98195, USA}
\begin{abstract}
\end{abstract}
\date{\today }

\maketitle
{\it The ``d-wave'' symmetry of the superconducting order in the cuprate high temperature superconductors is a well established fact \cite{dwave,dwave1}, and one which identifies them as ``unconventional.''
However, in macroscopic contexts -- including
many potential  applications ({\it i.e.} superconducting ``wires'') --  the material is a composite of
 randomly oriented superconducting grains in a metallic matrix, in which  Josephson coupling between grains  mediates the onset of long-range phase coherence.  (See,
 {\it e.g.}, \cite{Larbalestier,Malozemov,Heine}.) Here, we analyze the physics at length scales large compared to the size of such grains, and in particular the macroscopic character of the long-range order that emerges.  While XY-glass order and  macroscopic d-wave superconductivity %are
 may be  possible, we show that under
many circumstances -- especially when the d-wave superconducting grains are embedded in a metallic matrix -- the most likely order has global s-wave symmetry.}

\noindent{\bf Classification of phases:}

The anomalous average of the spin-singlet electron pair-creation operator which characterizes the superconducting state is
 \footnote{As is conventional, we will ignore the dynamical fluctuations of the electromagnetic gauge fields, as issues related to Elizer's theorem and gauge invariance are not important in the present context.\cite{shivajiandvadim}}
\begin{equation}\label{Defin}
\langle \phi(\vec r,\vec r^\prime)\rangle \equiv  \langle\left[\psi_{\uparrow}(\vec r)\psi_{\downarrow}(\vec r^\prime) + \psi_{\uparrow}(\vec r^\prime)\psi_{\downarrow}(\vec r)\right ]\rangle/\sqrt{2},
\end{equation}
where $\psi^\dagger_\sigma(\vec r)$ creates an electron with spin polarization $\sigma$ at position $\vec r$, and $\langle \ \rangle$ represents the equilibrium  average over all thermal and quantum fluctuations.

Above  the critical temperature, $T_{c}$,   in the normal metal phase, $\langle \phi(\vec r,\vec r^\prime)\rangle=0$,
while  $\langle \phi(\vec r,\vec r^\prime)\rangle\neq 0$ for all
 $T<T_{c}$.
In pure crystals this quantity is only a function of $(\vec r-\vec r')$, so it is convenient to introduce its Fourier transform $\langle \phi(\vec p)\rangle$.   Possible superconducting phases
were classified in Refs.~\cite{GorkovVolovik,Sigrist}.
  In particular, in s-wave superconductors,  $\langle \phi(\vec p)\rangle$ is invariant under all symmetry transformations of the crystal
  and $\langle \phi(\vec r,\vec r)\rangle \neq 0$, while
  for other forms of singlet order,
  $\langle \phi(\vec p)\rangle$
  changes sign under certain symmetry transformations, and consequently $\langle \phi(\vec r,\vec r)\rangle =0$. For example, in a d-wave superconductor, $\langle \phi(\vec p)\rangle$ (as well as $\langle \phi(\vec r,\vec r^\prime)\rangle$) changes sign under rotation by $\pi/2$,
  as is shown schematically in Fig. 2.

In a disordered system, $\langle \phi(\vec r,\vec r^\prime)\rangle$  is
a sample specific random quantity which does not posses any spacial symmetry.
As a result, in any singlet superconducting phase in a
disordered system, since no symmetry prevents it, generically $\langle \phi(\vec r,\vec r)\rangle \neq 0$.
\footnote{In the limit of infinite on-site repulsion, typically considered in studies of the ``$t-J$ model,''
there are local kinematic constraints which cause
$ \phi(\vec r,\vec r) $ to vanish when  $\vec r$ is taken to represent the location of a particular Wannier state, independent of the symmetry of the pairing.
 In such a case, care must be taken to properly define $\phi $ as a suitably course-grained quantity, but this does not affect any of our conclusions.}

In this article we assume that the disorder ensemble
preserves a group of translational and rotational symmetries statistically ({\it i.e.} on average).
 We will show below that after averaging over the realizations of the disorder, several superconducting phases emerge that can be precisely characterized by  different order parameters:

The internal symmetry of the
superconducting pairs (for example, s-wave or d-wave) refers to the transformation properties of the quantity
\begin{equation}
\Phi(\vec r-\vec r^\prime)=\overline{\langle\phi(\vec r,\vec r^\prime)\rangle}.
\label{simple}
\end{equation}
under rotation, where the overline indicates
a quantity that has been averaged over  configurations of the quenched variables,
{\it e.g.} the size, shape, orientation, and location of the superconducting grains as well as over realizations of the disordered potential in the metal.

Since the internal structure of $\Phi(\vec r)$  is not directly measurable, the  symmetry of the superconducting state is best {\it defined} in terms of ``phase sensitive measurements.''  Definitionally, this refers to any measurement of the relative phase of the order parameter at two macroscopically separated locations  on the surface of the system.  For instance, this can be measured in a SQUID  consisting of Josephson junctions at two surface positions  connected by  an external (macroscopic) conventional superconducting wire loop. In an s-wave state, the phase difference around the loop is  zero independent of the relative orientation of the two surfaces;  consequently in equilibrium there is no magnetic flux through the SQUID.  By contrast, in a d-wave state, a phase difference of $\pi$ is induced in a ``corner SQUID'' in which the loop connects portions of the surface approximately at right angles to each other, so there is a half quantum of magnetic flux through the SQUID in equilibrium\cite{dwave,dwave1}.  At the same time, if the SQUID loop connects nearly parallel portions of surface of a d-wave state (either on the same or opposite sides of the system), no  equilibrium flux is induced.  Similar analysis can be used to give a phase sensitive definition of other possible pairing symmetries.

The appropriate order parameter  characterization of
a superconducting glass state is a bit more subtle:
while $\langle\phi(\vec r,\vec r^\prime)\rangle\neq 0$, its local phase
varies randomly as a function of position, and correspondingly
 its configuration (or spatial) average vanishes.  However, this state is  sharply distinguished from the  normal state by the existence of a non-zero Edwards-Anderson type order parameter \cite{glass},
\begin{equation}
M(\vec r-\vec r^\prime)\equiv \overline{|\langle\phi(\vec r,\vec r^\prime)\rangle|^{2}},
\end{equation}
{\it i.e.} the glass state has $\overline{\langle\phi(\vec r,\vec r^\prime)\rangle}=0$ but $M\equiv M(\vec 0) >0$.

Another feature of
a glassy state, reflecting the existence of  random  variations of the phase of $\langle\phi(\vec r,\vec r^\prime)\rangle$, is that there are
local equilibrium currents, $ \langle \vec {\cal J}(\vec r)\rangle\neq \vec 0$, and associated spontaneous breaking of time-reversal symmetry.
It is easy to see that the configuration (or spatial) average current must vanish, $\overline{\langle \vec {\cal J}(\vec r)\rangle}=\vec 0$.
Instead, an appropriate  tensor
order parameter is
\begin{equation}
\tau_{ab} \equiv \overline{\langle  {\cal J}_a(\vec 0)\rangle\langle {\cal J}_b(\vec 0)\rangle}.
\end{equation}

A complete   classification of the various broken symmetry phases in disordered superconductors is not currently available.
However, in terms of the various quantities introduced above, we can
define the phases which we will encounter in our discussion:

\begin{itemize}

\item{1)} In the ``normal'' state at elevated temperatures, no symmetries are broken, and hence all the order parameters vanish.

\item{2)} In a ``s-wave superconducting'' state, $\Phi(\vec r)$ is non-zero, while $\tau_{a,b}=0$ -- {\it i.e.} there are no equilibrium currents in the bulk, and $\Phi(\vec r)$
is invariant under all symmetry transformations, {\it i.e.} it is rotationally invariant.  In phase sensitive measurements,
no equilibrium flux is induced in a SQUID, regardless of its geometry.

   \item{3)}  In a ``d-wave superconducting'' state, $\tau_{a,b}=0$, $\Phi(\vec r)$ is non-zero for non-zero $|\vec r|$ and changes sign under rotation by $\pi/2$ about some axis, but is invariant under rotation by $\pi$.
   Consequently, $\Phi(\vec 0)=0$ and a half flux quantum is induced in equilibrium in a suitable corner SQUID.

\item{4)} In a ``superconducting XY glass'' state, $\Phi(\vec r)$ vanishes but both $M$ and $\tau_{a,b}$ are non-zero.
 It is widely accepted, but still not completely settled that such a state exists in three dimensions below a non-zero  glass transition temperature.
There is also the possibility of a partially ordered glass phase with $M(\vec r)=0$ but $\tau_{a,b}\neq 0$, but whether this  arises in generic models is still being debated.\cite{Young,Kawamura}

    \item{5)}  It is possible to have a phase of coexisting uniform and glassy order -- in such a phase both $\Phi$ and $\tau_{ab}$ are non-zero.  Depending on the behavior of $\Phi(\vec r)$ under symmetry transformations (``rotations''), such a phase can still be classified as s-wave or d-wave etc.

\item{2b)} The most
unexpected new state we identify here is what we will henceforth refer to as a ``{\it globally s-wave superconducting}''  state.
Definitionally, this is   simply an unfamiliar limit of a  s-wave state -- one in which locally (i.e. in each superconducting grain) the order parameter has d-wave symmetry, but globally it is s-wave.  Spectroscopically, such a state  can reflect its microscopic origins as a d-wave superconductor, but from the viewpoint of macroscopic phase-sensitive measurements, it has s-wave symmetry.  Moreover, time-reversal symmetry is unbroken, $\tau_{ab} =0$.

\end{itemize}

\noindent{\bf The effective Hamiltonian:}
Below the bulk transition temperature, in a system composed of grains of size large compared to the coherence length, there is a  well developed magnitude of the order parameter on each grain.  We will consider the case
of most  relevance to the cuprates, in which the order parameter on each grain transforms according to a one-dimensional non-trivial (d-wave) representation of the point group.
The only important low energy degree of freedom is the overall phase of the superconducting order parameter on each grain, which we will designate $\theta_j$.

There is a degree of arbitrariness in the definition of $\theta_j$;   we choose a convention such that when $\theta_j=0$, the order parameter on the grain is real and  is positive in some particular crystalographic direction.
Moreover we assume that the grains are sufficiently large
  that we can neglect quantum fluctuations of the order parameter.  In this case, the macroscopic properties of the system can be captured by the phenomenological model
\begin{equation}
\label{JhosHamiltonian}
H= -\sum_{ij} J_{ij}\cos(\theta_i-\theta_j)
\ ,
\end{equation}
where the Josephson coupling $J_{ij}$ between grains $i$ and $j$ is real.

On general phenomenological grounds, we can separate the contributions to $J_{ij}$   into three  pieces,
\begin{equation}
\label{Eq:Jij}
J_{ij}= \eta_i\eta_j  J^{(1)}_{ij} + \eta_{ij}J^{(2)}_{ij} + J^{(3)}_{ij}.
\end{equation}
Here, $J^{(a)}_{ij}\geq 0$ for all $a$,
while $\eta_j=\pm 1$ and $\eta_{ij}=\pm 1$ are  random variables with vanishing mean which determine the sign of the corresponding contributions to $J_{ij}$.
The magnitude of each contribution to $J_{ij}$ is characterized by its mean, $\bar J^{(a)}\equiv \sum_j \overline{J^{(a)}_{ij}}$, and
to be explicit, we will always consider this model in three spatial dimensions, although it need not be  isotropic.

For grains of  conventional superconductors, under most circumstances, the Josephson coupling %is
would be positive, which is to say that $\bar J^{(1)}\approx \bar J^{(2)} \approx 0$;
 in this limit, the model is equivalent to an XY ferromagnet which has a single phase transition at $T_c\sim \bar J^{(3)}$, where physically the ``ferromagnetic phase'' corresponds to a statistically uniform s-wave superconducting phase.  (A mean-field estimate yields $T_c^{(MF)}=\bar J^{(3)}/2$.)

In any case in which the crystalline axes of d-wave superconducting grains are embedded into
a disordered metal with random orientations,
the sign of $J_{ij}$
is random, which means that $ \bar J^{(3)} =0$.
The two remaining terms in Eq~\ref{Eq:Jij}  have very different character:  The term proportional to $J^{(1)}_{ij}$ has its sign determined by a product of quantities that depend on the properties of each grain separately (which we will see is roughly related to the shape of the grains).
 Conversely, the sign of the term proportional to  $J^{(2)}_{ij}$ is determined by a joint property of the pair of grains (which we will see is related to the relative orientation of their crystalline axes).
 In the limit  $J^{(1)}=0$ (and $J^{(3)}=0$), this problem is a version of the standard model of an XY spin-glass,\cite{XYglass} while for $J^{(2)}=0$ (and $J^{(3)}=0$), this problem is a version of the well known Mattis model.\cite{Mattis}

Let us  redefine the zero of phase on each grain separately to introduce  ``Mattis-transformed'' phases
\begin{equation}
 \tilde \theta_j\equiv \theta_j +\pi(1-\eta_j)/2.
 \label{tildetheta}
 \end{equation}
 In terms of these transformed variables, the form of $H$ is unchanged, but the
 role of the different distributions is interchanged such that $\bar J^{(1)}\to \bar J^{(3)}$, $\bar J^{(2)}\to \bar J^{(2)}$, and $\bar J^{(3)}\to J^{(1)}$.   That the pure Mattis model is transformed in this way into a pure ferromagnetic XY model reflects the well known fact that this
 model introduces disorder without frustration.

\noindent{\bf %A c
Conjectured phase diagrams:}  In Fig. 1a we show a conjectured phase diagram for
the model
in Eq. \ref{JhosHamiltonian}  under the conditions that $\bar J^{(3)}=0$ as a function of the  dimensionless  temperature, $T/\bar J$, and the relative magnitude of the Mattis and spin-glass type couplings, $0\leq \bar J^{(2)}/\bar J \leq 1$, where $\bar J \equiv \sqrt{[\bar J^{(1)}]^2+[\bar J^{(2)}]^2}$.
We have labeled the ordered state at small $\bar J^{(2)}/\bar J$ ``globally s-wave'' as in the present context,
as we will show below, this state -- in which the Mattis transformed phases are uniformly (``ferromagnetically'') ordered -- corresponds to the
state defined in 2b), above.  The ``XY glass phase'' is  defined in 4), above.  The intermediate state, in the present context, has coexsiting global s-wave and XY glass order, corresponding to 5), above.  Formally, via the transformation in Eq. \ref{tildetheta}, the same phase diagram applies to the problem (which has  been studied in the spin-glass literature\cite{XYspinglass2}) of an XY spin glass with
an excess of ferromagnetic interactions ($\bar J^{(1)}=0$, while $\bar J^{(2)}$ and $\bar J^{(3)}\neq 0$).

The arguments  leading to this phase diagram
along the edges are as follows:

	i) For $\bar J^{(2)}=0$ the Mattis transformed problem is equivalent to a XY ferromagnet with some randomness  in the magnitude of the exchange couplings.  Thus, we conclude that there is a single phase transition with $T_{c} \sim \bar J^{(1)}$ to a phase with long-range
 superconducting order.

ii)
Decades of work
has still not resulted in a well established understanding of even the most basic features of the XY spin-glass in $d=3$.
It is  well accepted that there is a thermodynamic transition to a XY glass phase in large enough $d$. (See for example Ref.~\cite{Young}.) However, while  it is widely believed that  this conclusion applies in $d=3$ (but not in $d=2$), there remains some uncertainty concerning this conclusion.\cite{Kawamura}.
 Numerical experiments  certainly reveal that if there is a transition at  $T_{glass} \propto \bar J^{(2)}$ , the proportionality constant
must be    small ({\it i.e.} the spin-glass transition temperature is
at least an order of magnitude smaller than for the corresponding model without frustration).
Another intensely debated issue  is whether
 there exists an intermediate partially ordered phase with $M=0$ but $\tau_{ab} \neq 0$.
We have drawn  the phase boundaries in the limit  $\bar J^{(1)}\to 0$ in Figs. 1a and b  under the assumption that there is a finite $T_{glass}$ in d=3.  If it turns out that $T_{glass}=0$, the lines marking the boundaries of the various glassy phases should be reinterpretted as crossover lines below which relaxation rates become extremely small.  Conversely, if there are two transitions, then the lines should be interpretted as the mean of the two transitions.
Even if there is a transition, it is not clear whether the spin-glass phase would have a non-zero critical current.
Moreover, in a glassy state,  thermodynamic and transport quantities (including any ``apparent critical current'')
would be time dependent.
This is very different from the "globally s-wave superconducting state" where the critical current is finite and time independent. %!!!
Approaching the glass phase from above ($T>T_{glass}$), the tendency to a state with non-zero $\tau_{ab}$ should give rise to a growing {\it paramagnetic} response -  related to the so-called ``Wohlleben effect.''\cite{wohlleben,SigristRice}

We can also analyze the effect of moving in slightly from the edges of the phase diagram.
Adding a small excess of ferromagnetic couplings in an XY spin-glass is not thought \cite{gingras} to fundamentally affect the nature of the glass phase, so one does not expect to encounter any new phases   moving in from the right edge of the phase diagram into the regime in which  $\bar J^{(2)}\gg \bar J^{(1)}>0$.  Similarly, at non-zero $T$,  the same is true near the left edge of the phase diagram where $\bar J^{(1)}\gg \bar J^{(2)}>0$ and $T >  \bar J^{(2)}$.

We note that strictly speaking, in the framework of Eq.~\ref{JhosHamiltonian},
the ``globally s-wave'' state exists only at non-vanishing temperatures.
To see this, we consider the problem in terms  of the Mattis transformed
variables, so we consider the situation at low temperature, deep in the uniformly (ferromagnetically) ordered phase.  Even though $\bar J^{(2)}\ll \bar J^{(1)}$, there is some non-vanishing probability (proportional to $\bar J^{(2)}/\bar J$ )
 to find some pairs of nearby grains coupled by  a strong negative (frustrating) Josephson coupling.  Consequently,  in the
 ground-state, there will be a small vortex loop enclosing this bond. It is easy to see that two such double-degenerate defects interact as dipoles, so they will in turn freeze into a dipolar glass phase at a temperature which is roughly proportional to their concentration, i.e. at least to first approximation $T_{glass} \sim [\bar J^{(2)}/\bar J^{(1)}]$.
 Thus, at low enough temperatures and non-zero $\bar J^{(2)}$ there occurs a phase with
 globally s-wave superconducting order coexisting with time-reversal symmetry breaking vortex glass order. At $T>T_{glass}$ the thermal fluctuations restore
 time reversal
 symmetry, leaving the globally s-wave state.

Quantum fluctuations of the order parameter, which are not included in the model in Eq.~\ref{JhosHamiltonian}, produce
effects similar to non-zero $T$;
quantum tunnelling between the two degenerate vortex loop states of an isolated  defect thus effectively restore  time reversal symmetry. This makes it possible for the globally s-wave state to exist for small enough  $\bar J^{(2)}$ even at $T=0$ as indicated by the dotted line in Fig. 1a.
It is not currently known how the superconducting phases merge at higher temperatures and intermediate $\bar{J}^{2}/\bar{J}^{1}$, and indeed it is likely non-universal.  We have shown with dashed lines one possible, particularly simple completion involving a single tetra-critical point.

In Fig. 1b, we show a similar phase diagram for the situation in which $\bar J^{(2)}=0$ while $\bar J^{(1)}$ and $\bar J^{(3)}$ are non-zero -- this is relevant to the case  in which the grains are all oriented, but have random shapes and separations.
The nature of the phase diagram along the ``edges'' follows from the same sort of analysis that led to Fig. 1a.
 Where the $\bar J^{(1)}\gg \bar J^{(3)}$ we find the by now familiar globally s-wave state, while for  $\bar J^{(3)}\gg \bar J^{(1)}$ we encounter the conventional d-wave state.  At low temperatures, reflecting the intrinsic frustration when both $\bar J^{(1)}$ and $ \bar J^{(3)}$ are comparable, both of these phases give way to a phase with coexisting glassy order (and hence subject to all the associated caveats).  The phase diagram possesses a reflection symmetry implied by the Mattis transformation in Eq. \ref{tildetheta}.
Again,
 even
  the topology of the middle regions of the phase diagram is
  likely non-universal;  with dashed lines we show a plausible minimal completion.
The dotted  lines indicate
 the region where (were we to include them in the model) quantum fluctuations of the order parameter destroy the glassy order of the vortex loops.

\begin{figure}[ptb]
\includegraphics[width=4.2cm]{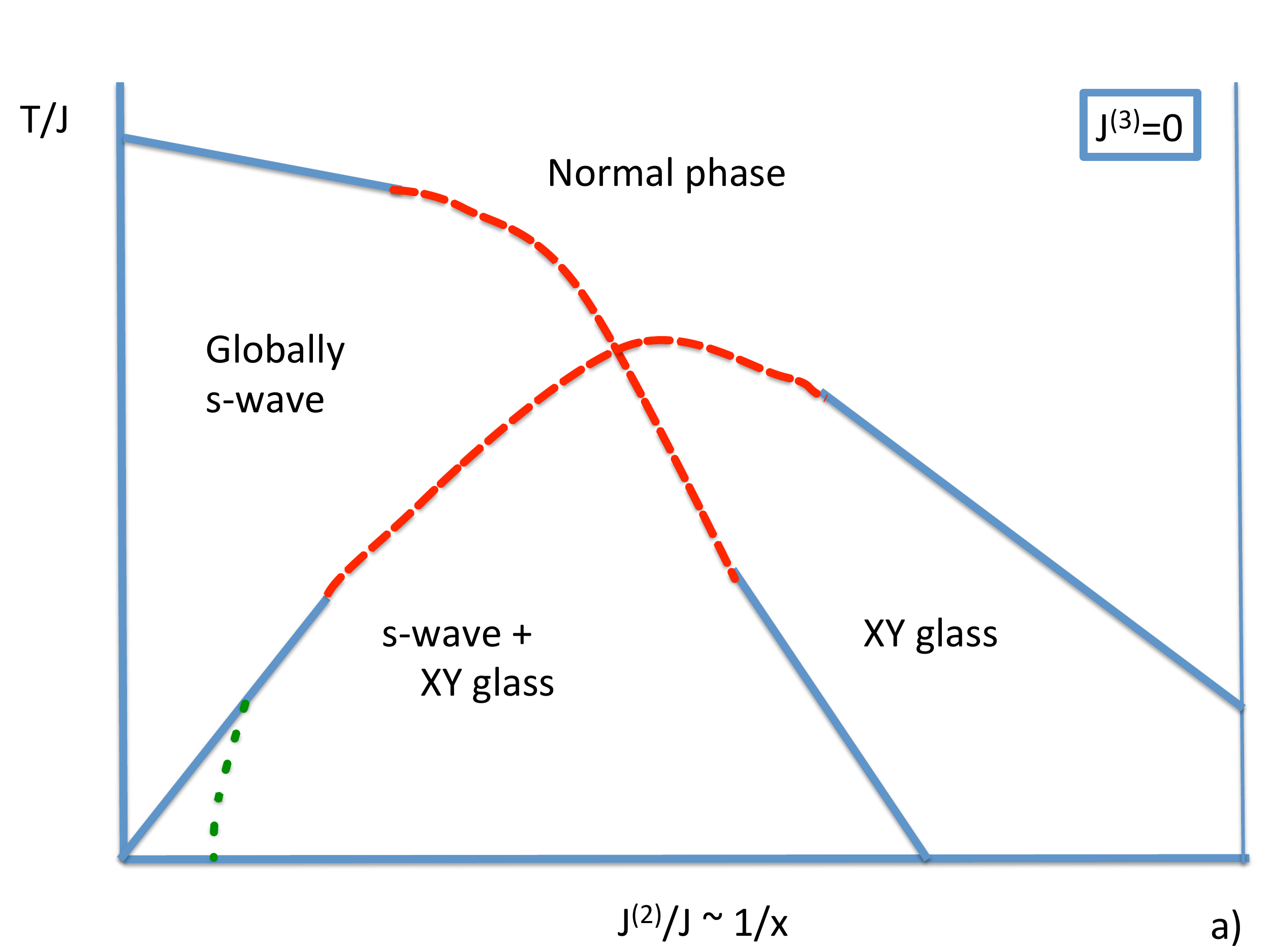}
\includegraphics[width=4.2cm]{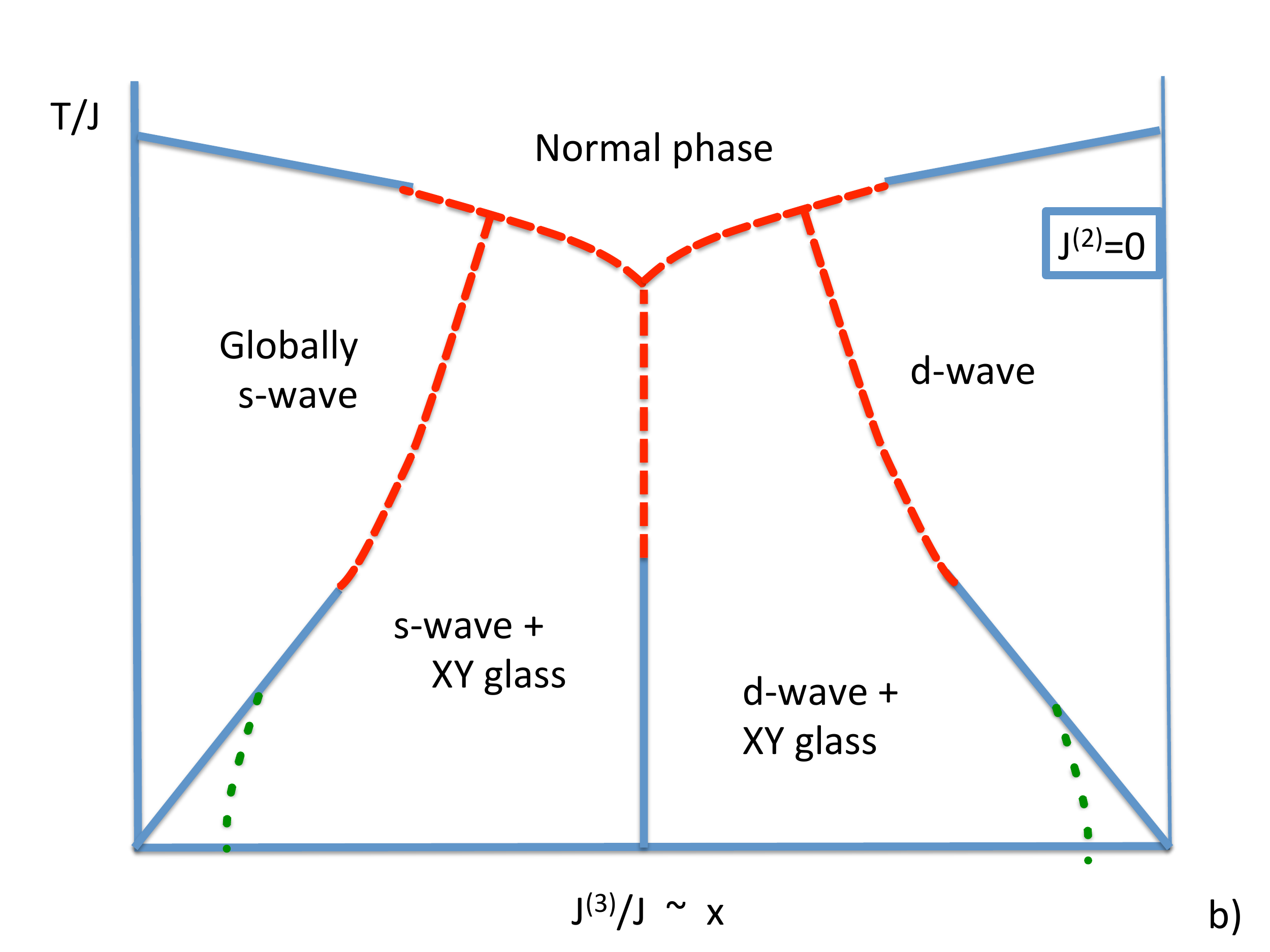}
\caption{Schematic phase diagrams for a) $\bar J^{(3)}=0$ and $J\equiv \sqrt{(\bar J^{(1)})^2+(\bar J^{(2)})^2}$ and b) $\bar J^{(2)}=0$ and $J\equiv \sqrt{(\bar J^{(1)})^2+(\bar J^{(3)})^2}$.  The qualitative structure of the solid lines is justified by asymptotic analysis in the text in the context of the (%still controversial
unproven) assumption that there is a finite spin-glass transition temperature for the XY spin glass in d=3, while the dashed lines are included as a representative guess of how the phase diagram might be completed.  The dotted lines represent the leading effect expected from quantum fluctuations -- which are not explicitly included in the model.} \label{fig:fig1}
\end{figure}

\noindent{\bf Proximity effect and the Josephson couplings:}
We now
address the origin  of the various terms in Eq.~\ref{Eq:Jij}.
 The Josephson coupling between grains of a d-wave superconductor separated by  normal metal is controlled by the proximity effect.  Andreev reflection at the superconductor-normal metal boundary generates a finite value $\langle \phi(\vec r,\vec r^\prime)\rangle$ inside the normal metal.
 We will consider the case where the grains
 are large compared to  both the coherence length, $\xi_0$, and the
 mean free path $l$.

We begin by considering a single isolated grain embedded in a disordered normal metal matrix. It is important to emphasize that, %even if
 whether or not the order parameter inside the grain
 has d-wave character, the superconductor-metal interface
 breaks the point-group symmetry
 so
 a finite value of $\langle \phi(\vec r,\vec r)\rangle$ is induced in the neighboring metal. We will call the quantity $\overline{\langle \phi(\vec r,\vec r)\rangle}$ the local s-wave component.
  It is this component which survives in the normal metal on distances larger than $l$, while the local d-wave and other-wave components
  decay exponentially at large distances.

      To obtain the requisite boundary condition
      for $\overline{\langle \phi(\vec r,\vec r)\rangle}$ at the normal metal-superconductor
      interface, one has to match its values in the superconductor and in the normal metal close to the
      interface.
   In particular, its sign reflects the sign of $\langle \phi(\vec p_\perp)\rangle$ in the grain's bulk
   for $\vec p=\vec p_\perp$ normal to the
   interface ($|p|=p_{F}$).
   Thus, the sign of $\overline{\langle \phi(\vec r,\vec r)\rangle}$   changes along the boundary.

\begin{figure}[ptb]
\includegraphics[width=8.0cm]{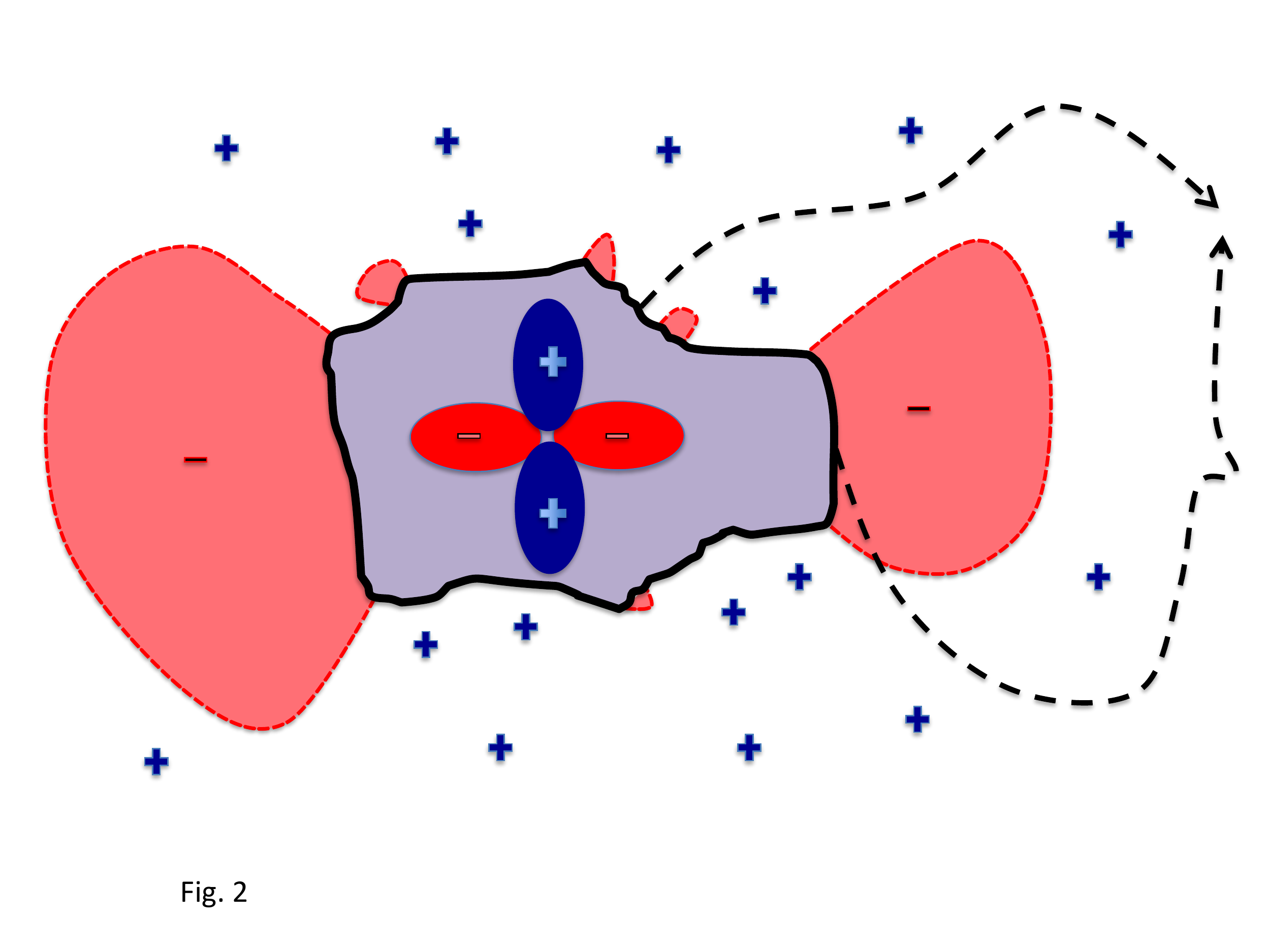}
\caption{Schematic representation of the sign of the anomalous average of the pair creation operator produced in the surrounding disordered metal by the proximity effect  coupling to a grain of a d-wave superconductor. The symbol inside the grain represents the structure of $\langle\phi(\vec p)\rangle$ inside the grain.  The dashed lines represent typical pair diffusion paths the contribute to the proximity effect. } \label{fig:fig2}
\end{figure}

   The key
   consequence is illustrated graphically in Fig.~\ref{fig:fig2}, which is a schematic of a grain of a d-wave superconductor embedded in a metallic matrix. At distances from the grain larger than $l$ but small compared to the size of the grain, %it
   the sign of  $\overline{\langle \phi(\vec r,\vec r)\rangle}$
 is dominated by the
 closest interface, and so it
%which
can be positive or negative depending on the orientation
of that interface.
However, far from the grain,
the value of $\overline{\langle \phi(\vec r,\vec r)\rangle}$ gets contributions from Cooper pairs that
have scattered from all parts of the grain's surface  and then diffused to the observation point $\vec r$. Thus its sign is determined by the sign of $\overline{\langle \phi(\vec r,\vec r)\rangle}|_{S}$
averaged over the entire surface of the grain.
In other words {\it at large distances, the sign of the anomalous average is fully determined by the geometry of the grain.}
  On the qualitative level the sign of $\overline{\langle \phi(\vec r,\vec r)\rangle}$ is determined by the interference between the Cooper pairs traveling to the observation point $\vec r$ along different diffusive trajectories,
  as illustrated  by the two representative paths shown as the dashed lines in Fig.~\ref{fig:fig2}.

 To compute the Josephson coupling between a pair of grains formally requires the solution of this  problem  in the presence of two superconducting grains as a function of their   relative phases.  However,
it  is straightforward to  understand the sign of the Josephson coupling
 by overlaying the patterns $\langle \phi(\vec r,\vec r)\rangle$ produced by
 each grain individually.
As a result,
 for a pair of grains of comparable characteristic size $R$ separated by a distance $L$, we
 find that for $L> R$, the sign of $J_{ij}$ is determined by a product single grain characteristics \cite{oreto,oreto1,Andreev} -- {\it i.e.} it appears as a contribution to $J^{(1)}_{ij}$ -- while for $L< R$, its sign is determined by mutual aspects of the shape and orientation of the two grains -- {\it i.e.} it appears as a contribution to $J^{(2)}_{ij}$.  Other than this, the considerations controlling of the magnitude of the Josephson coupling are not substantially different from those in  conventional S-N-S junctions.  For instance, for $L\gg R\gg l$
 \begin{equation}\label{JijFormula}
%J_{ij}^{(1)}\sim B\frac{R^{3}}{|\vec r|^{3}}e^{-|\vec r|/]\xi_T}
J_{ij} = \eta_i\eta_j J_{ij}^{(1)}\sim \eta_i\eta_j\left(\frac{GDR}{|\vec r|^{3}}\right)e^{-|\vec r|/]\xi_T}
\end{equation}
where $D$ is the electron diffusion coefficient and
$\xi_T = \sqrt{D/T}$  the coherence length of the surrounding metal,  and $G$ is the characteristic conductance of the grain.
In $d=3$,  generally  $G\propto R$ so the term in parentheses scales as
 $J_{ij} \sim R^2/L^3$.

Importantly for present purposes, pairs of randomly oriented grains with $R \gtrsim L$ contribute to $J^{(2)}$ while those with $L\gtrsim R$ contribute to $J^{(1)}$.  Consequently, for fixed size grains, the material can be tuned across the phase diagram in Figs.~\ref{fig:fig1}a \& b by varying the concentration of grains, as indicated.

{\bf Identification of  ``globally s-wave'' order:}  It should now be apparent that the global s-wave symmetry of the superconducting order
refers to the symmetry of the
induced order in the metallic host produced by the embedded d-wave grains.  Even though the superconductivity originates within the grains, the relative phase of order parameter from grain to grain is determined by the condition that the phase be constant throughout the metal.
Another way to obtain an intuitive understanding is to imagine
replacing the metal by an s-wave superconductor.  Now, consider the effect this has on the phase of the superconducting order in each (d-wave superconducting) grain.  It is clear that in this case, baring an accidental degeneracy, the phase of the d-wave order parameter on each grain will be locked to that of the surrounding metal -- either with the same phase or with a phase-shift of $\pi$ depending on the shape of the grain.
If we now imagine continuously decreasing the strength of the intrinsic s-wave order in the matrix, by adiabatic continuity we would approach the situation we have discussed here.

\noindent{\bf Further implications:}  In addition to its fundamental interest, the present results suggest
 new strategies for making better practical wires.  Firstly, to avoid the various detrimental effects of frustration (and glassy phases), one would like to insure that $\bar J^{(1)}>\bar J^{(2)}$;  this is accomplished  by insuring that the separation, $L$, between the grains is larger than or comparable to their characteristic radius, $R$.
 However, it is also desirable that the magnitude of the Josephson couplings be as large as possible.  At the very least, this implies that we would like $L < \xi_T$ corresponding to the temperature (less than the bulk $T_c$) at which the wires are to be used.  Moreover, even when this inequality is satisfied, the coupling between two neighboring grains scales as $J\sim R^2/L^3$.  These various different considerations suggest that improved wires can be obtained by reducing the grain size and simultaneously increasing the concentration of grains subject to the condition $L/R \gtrsim 1$.  In particular,
 there are likely regimes in which increasing $L$ causes a transition from the spin-glass  to the globally s-wave regime, and correspondingly an increasing magnitude and decreasing time dependence of the critical current.

Of course, this strategy has its limits -- in order that the grains have undiminished local superconducting order, it is necessary that $R \gg \xi_0$,
and that
$R$ is  large enough that quantum fluctuations of the superconducting phase are negligible.
Subject to this, for the cuprates, it would be particularly interesting to explore the situation in which the London penetration depth, $\lambda$, is large compared to the grain size, $\lambda > L \sim R \gg \xi_0$, in which case the superfluid density of the wire would be homogeneous and isotropic, thus potentially mitigating some of the undesirable consequences of the quasi-2d nature of the cuprates.\cite{beasley}

\acknowledgements{We would like to acknowledge useful discussions with M. Beasley, M. Gingras, D. Huse, K. Moler and A. P. Young. This research was supported in part by Department of
Energy under grant number DE-AC02-76SF00515 at Stanford (S.A.K.)}

\end{document}